\documentclass[a4paper,showkeys,nofootinbib]{revtex4}
\usepackage[utf8]{inputenc}
\usepackage{graphicx,bm,color}
\usepackage{amsmath}
\usepackage{slashed,verbatim}
\usepackage{subfigure}
\usepackage[linktocpage=true]{hyperref}
\usepackage{gensymb}

\begin{document}
                                                                                                                                                                                                                     
\title[Constraints on dark matter models from the observations of Triangulum-II with the \textit{Fermi} Large Area Telescope]{
Constraints on dark matter models from the observation of Triangulum-II with the \textit{Fermi} Large Area Telescope }  

\author {Sayan Biswas$^{1}$}
\email []{sayan@rri.res.in}
\author {Pooja Bhattacharjee$^{2,3}$}
\email []{pooja.bhattacharjee@jcbose.ac.in}
\author {Pratik Majumdar$^{4}$} 
\email []{pratik.majumdar@saha.ac.in}
\author {Subinoy Das$^{5}$} 
\email []{subinoy@iiap.res.in}
\author {Mousumi Das$^{5}$}
\email []{mousumi@iiap.res.in}
\author {Partha S.~Joarder$^{2,3}$}
\email []{partha@jcbose.ac.in} 
\affiliation{ $^{1}$ Raman Research Institute, C.V. Raman Avenue, Sadashivanagar, Bangalore, Karnataka  560080, India}
\affiliation{ $^{2}$ Department of Physics, Bose Institute, 93/1 A.P.C. Road, Kolkata 700009, India} 
\affiliation{ $^{3}$ Centre for Astroparticle Physics and Space Science, Bose Institute, Block EN, Sector V, Salt Lake, Kolkata 700091, India}
\affiliation{ $^{4}$ Saha Institute of Nuclear Physics, HBNI, 1/AF, Bidhannagar, Kolkata 700064, India}
\affiliation{$^{5}$ Indian Institute of Astrophysics, Koramangala, Bangalore, Karnataka 560034, India}


\begin{abstract}
\noindent Triangulum-II, a newly discovered dwarf spheroidal galaxy, is a strong candidate for the indirect search of dark matter through the detection of $\gamma$-ray emission that could originate from the pair- annihilation of
weakly interacting massive particles (WIMPs). We here report on the analysis of almost seven years Fermi Gamma-Ray Space Telescope data of Triangulum-II which was taken during its all sky survey operation mode. No excess $\gamma$-ray emission has been detected above 100 MeV from Triangulum-II. We derive the upper limits on $\gamma$-ray flux assuming both the power-law spectra and the spectra related to WIMP annihilation. In this work, we have considered several theoretical WIMP (neutralinos here) models envisioning both thermal and non-thermal production of WIMPs, and put limits on pair-annihilation cross-section of WIMPs to constrain the parameter space related to those
theoretical models.

\keywords{dark matter, WIMP, dwarf spheroidal galaxy.}
\end{abstract}
\maketitle

\section{Introduction}
\noindent Recent high precision data on the cosmic microwave background (CMB) \cite{bib:kom,bib:ade} have indicated that some form of non-luminous and non-baryonic matter, known as the dark matter (DM), may constitute around $75\%$ of the total mass density of the universe. In this context, the cosmological simulations and the theoretical arguments in recent times mostly favour the existence of certain cold dark matter (CDM) candidates to explain the formation of the observed large-scale structure of the universe \cite{bib:abdo}. In this scenario, a galactic halo, consisting of the CDM material, envelops the galactic disk that extends well beyond the visible edge of the galaxy. The actual constituents of such CDM halo, however, remains hitherto undetected in direct experiments and observations. On theoretical grounds, it is now believed \cite{bib:abdo,bib:ste,bib:jung} that a new type of weakly interacting massive particles (WIMPs) may be the most promising candidate for the constituent of such CDM. This is because of the fact that the relic thermal abundances of those WIMPs, as inferred from their plausible pair-annihilation cross sections (with their natural masses lying in the range from a few GeV to a few TeV), are comparable to  their abundances estimated from the cosmological DM calculations. It is believed that the pair-annihilation (or, the decay) of the WIMPs, that takes place even in the present epoch, is likely to yield high-energy $\gamma$-rays \cite{bib:abdo}. The detection of such high-energy $\gamma$-rays in the galactic halos is, therefore, expected to provide us with some indirect information regarding the signatures of the CDM. In this paper, we assume that an individual WIMP particle is intrinsically stable, so that, only the pair-annihilations of such WIMPs has to be considered for the purpose of analysing the high-energy $\gamma$-ray data collected by the \textit{Fermi} Large area Telescope\footnote{\url{http://fermi.gsfc.nasa.gov}} ({\em {Fermi}}-LAT; referred sometimes simply as the `LAT' in this paper).

According to the cosmological N-body simulations of the structure formation, the CDM halos formed by the WIMPs are not smooth. On the contrary, such halos contain a large number of bound substructures (i.e., the sub-halos), whose number increases with decreasing mass of the WIMPs \cite{bib:die,bib:kuh,bib:spr,bib:abdo}. Those smaller halos (or, the sub-halos) in a large galactic halo are likely to appear as the so-called dwarf spheroidal galaxies (dSphs) in the actual observations. According to the CDM hypothesis, those dSphs may be considered to be the densest DM regions in the galactic halo; they are predicted to be the largest galactic substructures around the Milky Way Galaxy. The values of the mass-to-light ratio of such dSphs are likely to be in the range $(100-1000)~M_{\odot}/L_{\odot}$, in which $M_{\odot}$ and $L_{\odot}$ are the solar mass and the solar luminosity, respectively. The above might imply that the dSphs could mostly be the DM dominated structures in the halo of our galaxy. Lying away from the galactic centre, which produces the strongest and the most poorly understood hard gamma rays from myriad of physical sources, those dSphs could be the ideal sites for an indirect search of DM through the detection of the signals of WIMP annihilation that are less likely to be contaminated by the conventional backgrounds of baryonic origin. 

In the recent past, the Sloan Digital Sky Survey (SDSS) \cite{bib:york} discovered a new population of Milky Way satellites, the members of which are ultra-faint,  thus leading to the possibility that they could be the DM dominated dSphs \cite{bib:wil,bib:zuc,bib:bel,bib:irw,bib:wal,bib:str}. Indeed, over the past year or so, the Panoramic Survey Telescope and Rapid Response System (Pan-STARRS) \cite{bib:kai} and the Dark Energy Survey (DES) \cite{bib:abb} have found new candidate dSphs \cite{bib:lae,bib:bec,bib:kim,bib:drl} in the vicinity of the Milky Way. Triangulum-II (hereafter referred to as Tri-II), which is one of those newly discovered dSphs, has recently been closely investigated by the Pan-STARRS Survey \cite{bib:lae}. This survey has concluded that Tri-II is either an ultra-faint and DM dominated dwarf galaxy or a globular cluster. A number of calculations \cite{bib:gen,bib:hay} have also claimed that Tri-II may indeed be a potential target to search for the signatures of pair-annihilation of the WIMPs. The present investigation of Tri-II, that is described in this paper, is motivated by the the above findings.

The paper is organised in the following lines. In Section~2, we briefly summarise the properties of Tri-II. There, we also describe the procedure for the analysis of the \textit{Fermi} data on Tri-II as well as the method, in which we obtain the upper limit of the $\gamma$-ray fluxes from Tri-II by employing power-law spectra of different spectral indices. In the next sub-section (Section~3.1), we model the possible DM mass density in Tri-II by using a suitable density profile. In this sub-section, we also describe the parameters required to obtain the possible $\gamma$-ray flux arising from the pair-annihilation of the WIMPs constituting the DM in Tri-II. In the sub-section~3.2, we determine the upper limit of the possible $\gamma$-ray fluxes from Tri-II by using the spectra resulting from the annihilation of the WIMPs of different masses within the framework of various WIMP models. In that sub-section, we also determine the possible upper limits of the pair-annihilation cross-sections of the WIMPs and then discuss the implication of such calculations in the context of different WIMP models (Section~3.2). The conclusions of the paper are finally summarised in Section~4.

\section{\textit{Fermi}-LAT observation and data analysis of Tri-II}
\subsection{Tri-II}

In this paper, we confine ourselves to one of the prevailing models of Tri-II that considers it to be a metal-poor galaxy with rather large a mass to light ratio, but containing only a handful of member stars, the exact number of which is yet uncertain \cite{bib:kir,bib:kir1}. Earlier observations \cite{bib:kir, bib:mar1} had predicted some $6$ to $13$ member stars in Tri-II, while a recent study \cite{bib:kir1} seems to have confirmed the existence of $13$ stars with their velocity dispersion $\sigma_{\rm v}<4.2~{\rm {km}}~{\rm s}^{-1}$ and $<3.4~{\rm {km}}~{\rm s}^{-1}$; the confidence levels (C.L.) of those measurements being $95\%$ and $90\%$, respectively. Some of the important properties of Tri-II, that have, so far, been suggested by the observations, are summarised in Table~I.

\begin{table}[h!]
\begin{center}
\caption{Some properties of Tri-II}
\begin{tabular}{|p{4 cm}|p{4 cm}|p{2 cm}|}
\hline
\hline
Property &  Value   & Reference  \\ 

\hline

Galactic latitude & $\rm{141.4^{\circ}}$  & \cite{bib:lae} \\
\hline
Galactic longitude & $\rm{-23.4^{\circ}}$ & \cite{bib:lae} \\
\hline
Galactocentric distance & $\rm{36_{-2}^{+2}~kpc}$ & \cite{bib:gen,bib:kir}\\
\hline
2D half light radius ($\rm{r_{h}}$) & $\rm{34_{-8}^{+9}~pc}$ & \cite{bib:kir,bib:kir1} \\
\hline
Velocity relative to galactic standard of rest (GSR) ($\rm{v_{GSR}}$) & -261.7 km $\rm{s^{-1}}$ & \cite{bib:kir1}\\
\hline
Mean heliocentric velocity  $~<\rm{v_{helio}}>$ & $\rm{-381.7\pm2.9~km~s^{-1}}$ & \cite{bib:kir1}\\
\hline
Stellar Velocity Dispersion ($\rm{\sigma_{v}}$) & $<~\rm{3.4~km~s^{-1}~(90\%~C.L.)}$ & \cite{bib:kir1}\\
 & $<~\rm{4.2~km~s^{-1}~(95\%~C.L.)}$ & \cite{bib:kir1}\\
\hline
Mass within 3D half-light radius \Big($\rm{\frac{M_{1/2}}{M_{\odot}}}$\Big) & $\rm{<~3.7~\times~10^{5}~(90\%~C.L.)}$ & \cite{bib:kir1}\\
& $\rm{<~5.6~\times~10^{5}~(95\%~C.L.)}$ & \cite{bib:kir1}\\
\hline
Mass-to-light ratio within 3D half-light radius \Big($\rm{(M/L_{v})_{1/2}}$\Big) & $\rm{<~1640~M_{\odot}~L_{\odot}^{-1}~(90\%~C.L.)}$ & \cite{bib:kir1}\\
& $\rm{<~2510~M_{\odot}~L_{\odot}^{-1}~(95\%~C.L.)}$ & \cite{bib:kir1}\\
\hline
Density within 3D half-light radius $\rm{\rho_{1/2}}$ & $\rm{<~2.2~M_{\odot}~pc^{-3}~(90\%~C.L.)}$ & \cite{bib:kir1}\\
& $\rm{<~3.3~M_{\odot}~pc^{-3}~(95\%~C.L.)}$ & \cite{bib:kir1}\\
\hline
Metallicity ([$\rm{Fe/H}$]) & $\rm{-2.24\pm0.05}$ & \cite{bib:kir1}\\

\hline
\hline
\end{tabular}
\end{center}
\end{table}
\begin{center}
\end{center}

In Table~I, the quantities $M_{\odot}$ and $L_{\odot}$ denote the mass and the bolometric luminosity of the Sun, respectively. In this table, the quantities $\rm{M_{1/2}}$, $\rm{(M/L_{v})_{1/2}}$ and $\rm{\rho_{1/2}}$ have been determined~\cite{bib:kir, bib:kir1} by assuming Tri-II to be a spherically symmetric object in the state of dynamical equilibrium. A number of observational characteristics of Tri-II, namely, its large velocity with respect to the galactic standard of rest (GSR), its low ellipticity, the near-Gaussian nature of the observed line of sight velocity distribution of its member stars, and the rather large a value of its tidal radius (measured with respect to the Milky Way Galaxy) in comparison with the value of its 3D half-light radius, have now been revealed. All those observations have led us to believe that Tri-II has not, so far, been considerably affected by the total tidal effect from the Local Group of galaxies~\cite{bib:kir} including the Milky Way Galaxy. Also, the so-called associations of Tri-II with the Triangulum-Andromeda halo sub-structures \cite{bib:lae,bib:maj} and with the PAndAS stream \cite{bib:mar2}, that were earlier believed to be the signatures of tidal disruption of Tri-II, have now been completely ruled out \cite{bib:dea} on the ground of the order of magnitudically smaller GSR velocities of those substructures in comparison with the one pertaining to Tri-II. Admittedly, the above observations cannot provide a concrete proof in support of a state of dynamical equilibrium of Tri-II. Those observations, nevertheless, indicate towards the existence of rather weak a tidal effect of the Local Group on Tri-II \cite{bib:kir}. In this paper, we have assumed Tri-II to be in dynamical equilibrium with the {\em {caveat}} that a conclusive evidence to settle this particular issue, in one way or the other, is as yet unavailable.

\subsection{The \textit{Fermi}-LAT data analysis of Tri-II}

The \textit{Fermi}-LAT is a space-based $\gamma$-ray detector launched on June 11, 2008 by the Delta~II Heavy launch vehicle. It detects the $\gamma$-photons with their energies ranging from about $20$ MeV to about $300$ GeV \cite{bib:atw}. In our analysis, we have used the data accumulated over almost seven years (i.e., from August 4, 2008 to May 22, 2015) of observations of Tri-II during an all-sky survey operation mode of the above detector. 

In our analysis of the $\gamma$-ray data from Tri-II, we have used the version \texttt{v10r0p5} (released on June~24, 2015) of the software package \texttt{Fermi ScienceTools}\footnote{\url{https://fermi.gsfc.nasa.gov/ssc/data/analysis/software/}} (referred hereafter simply as the \texttt{ScienceTools}), that is dedicated for analysing the LAT-data. Here, we use the recently released, fully reprocessed Pass8 dataset\footnote{\url{https://fermi.gsfc.nasa.gov/ssc/data/analysis/documentation/Pass8_usage.html}}, that provides an improved event reconstruction, a wider energy range, a better measurement of the (reconstructed) energies and a significantly increased effective area, especially in the low energy range. As the region of interest (ROI), we have chosen a region covering $10^{\degree}$ radius centred on Tri-II by the use of the `gtselect'\footnote{\url{https://fermi.gsfc.nasa.gov/ssc/data/analysis/scitools/data_preparation.html}} option of the \texttt{Science Tools}. By using the same tool, we also apply a cut $0.1 \le E \le 50$~GeV on the reconstructed energy ($E$) of the photon events. This range is chosen to avoid calibration uncertainties at low energy and background contamination at high energy. The albedo contamination is avoided by rejecting the events with their zenith angles satisfying $\theta < 90^{\degree}$, and also by selecting the good time intervals (GTIs) by using the `gtmktime' filter~\cite{bib:abdo} suggested in the \texttt{ScienceTools}. Next, we analyse the dataset by using the `binned likelihood technique'\footnote{\url{https://fermi.gsfc.nasa.gov/ssc/data/analysis/scitools/binned_likelihood_tutorial.html}}, as implemented in the \texttt{ScienceTools} \cite{bib:cas,bib:matt}. In this analysis, we have used the photon events P8R2{\_}SOURCE of the event class 128 (providing good sensitivity to the point sources and the moderately extended sources) and the photon-to-($e^{+}$,~$e^{-}$)~pair conversion type 3, in which the pair conversion is supposed to take place both at the FRONT and the BACK tracker-layers of the LAT, so that, the overall LAT instrument has a good point spread function (PSF) at low energy, simultaneously presenting a large effective area~\cite{bib:abdo} at high energy. While selecting the photon events as above, we adopt an `instrument response function (IRF)'\footnote{\url{https://fermi.gsfc.nasa.gov/ssc/data/analysis/documentation/Cicerone/Cicerone_LAT_IRFs/IRF_overview.html}} suggested in version P8R2\_SOURCE\_V6 of the appropriate manual \footnote{\url{https://fermi.gsfc.nasa.gov/ssc/data/analysis/documentation/Cicerone/}} of the \texttt{ScienceTools}. 

With the ROI chosen as above, our source model includes Tri-II, along with all the sources enlisted in the \textit{Fermi} 3FGL catalogue \cite{bib:ace}, that lie  within this ROI. Due to the lack of a pre-existing study of Tri-II by the {\em {Fermi}}-collaboration, an initial approximation would be to model Tri-II as a point source with a power-law spectrum. The spectral functional forms of all the other sources in the ROI have already been provided in the catalogue. In addition, the standard models of the galactic diffuse emission and its possible isotropic component (containing the possible extragalactic diffuse emission and the residual charged background contamination), currently used by the LAT-collaboration \footnote{\url{https://fermi.gsfc.nasa.gov/ssc/data/access/lat/BackgroundModels.html}}, are included by us in the background model of the analysis. To take the uncertainties in modelling those diffuse components into account, the independent normalisations of those diffuse components are kept free during our maximum likelihood fitting procedure. In that procedure, the spectral normalisation parameters for the sources within $5^{\degree}$ from Tri-II are left free, while the parameters of other sources in the ROI are kept fixed to their (3FGL) catalogue values. Moreover, the localisation of Tri-II is kept fixed during the likelihood fitting procedure. The later assumption is reasonable in view of the limited angular resolution of the LAT and the limited statistics available from Tri-II. In the following subsection, we have fitted the source spectrum of Tri-II, obtained from the LAT-data, in terms of the model-independent power-law spectra of various spectral indices.

\subsection{Results from the power-law modelling} 

   \begin{figure}
\subfigure[]
 { \includegraphics[width=0.6\linewidth]{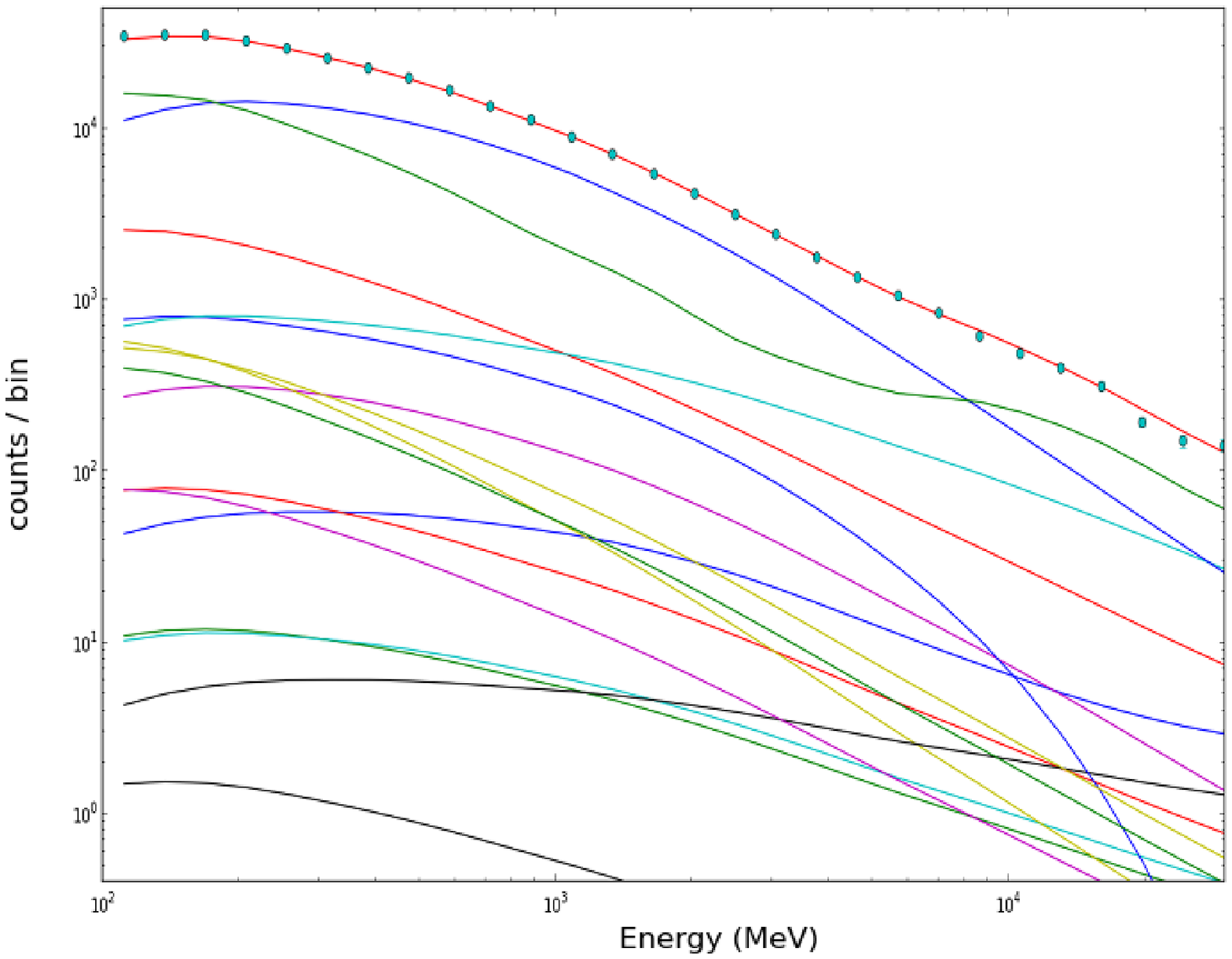}}
\subfigure[]
 { \includegraphics[width=0.6\linewidth]{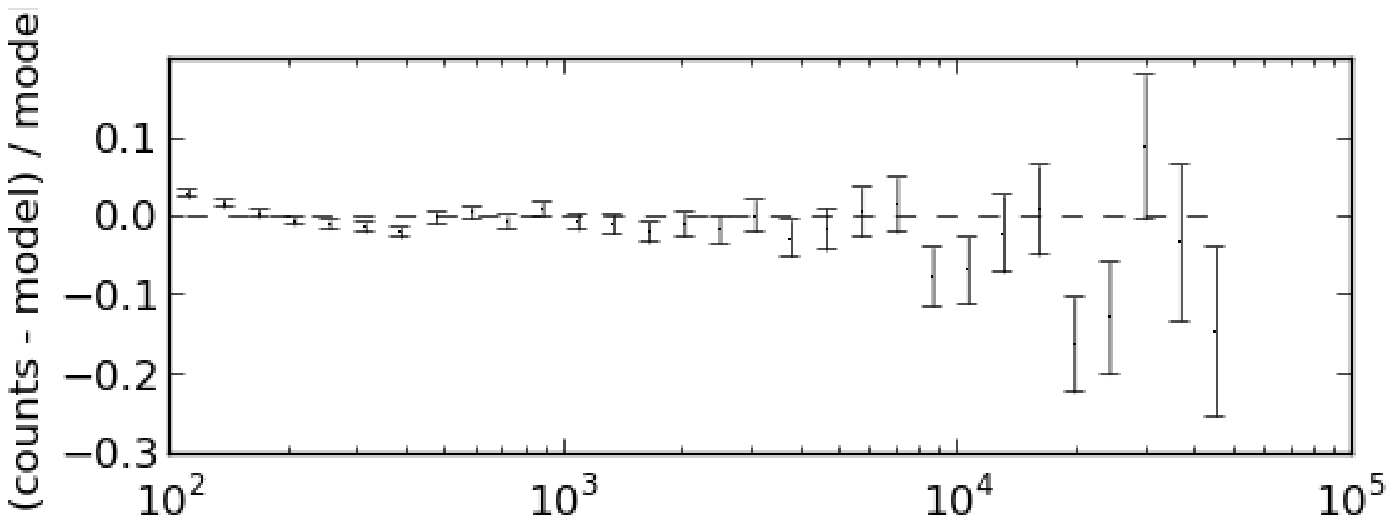}}
\caption{ Spectral fit to the counts (Fig.~1(a)) and the corresponding residual plot (Fig.~1(b)) for a chosen ROI centred on Tri-II. In Fig.~1(a), the power-law spectral index of Tri-II is chosen to be $\Gamma = 2$. In the same figure, the top red line displays the best-fit total spectrum, along with the corresponding LAT-observed data points; the top blue and the top green lines display the galactic diffuse component and the isotropic component, respectively. The rest of the coloured lines in Fig.~1(a) are for various point sources lying within the chosen ROI.}
\end{figure}

The differential photon flux, obtained from the observation of Tri-II by the {\em {Fermi}}-LAT after following the above analysis, is modelled by a power-law spectrum of the form~\cite{bib:abdo}:
\begin{equation}
\rm{\frac{dN}{dA dE dt} = N_{0} \Big(\frac{E}{E_{0}}\Big)^{-\Gamma}},
\end{equation} 
in which $dN$ is the number of photons, with their reconstructed energies lying within the interval from $E$ to $E + dE$ subject to the condition $0.1~{\rm {GeV}} \le (E, E+dE) \le 50~{\rm {GeV}}$, that are incident on an elemental area $dA$ of the detector in an elemental time interval $dt$.  In Eq.~(1), the spectral index $\Gamma$ is a variable parameter; $N_{0}$ is a normalisation constant to be determined by following the fitting procedure. Here, the arbitrary energy scale $E_{0}$ is set at $100~\rm{MeV}$, as appropriate for the energy interval considered in this work~\cite{bib:abdo}. In the modelling of the source spectrum, we consider five different values of the spectral index, namely, $\Gamma = 1$, $1.8$, $2$, $2.2$ and $2.4$, respectively~\cite{bib:abdo}. Here, $\Gamma = 1$ is inspired by the DM annihilation model found in Ref.~\cite{bib:ess}, while the last four indices put constraints on the standard astrophysical source spectra.

In our power-law modelling of the observed source spectrum, we have repeated the binned likelihood analysis for each of the above-mentioned spectral indices of Tri-II to determine the corresponding best-fit values of $N_{0}$ along with the isotropic and the galactic diffuse normalisations. Fig.~1(a) displays a sample of the results of such spectral fits to the data from all the sources, along with the diffuse and the isotropic background model, within the ROI considered in this subsection. Fig.~1(b), on the other hand, displays the residual of the fit displayed in Fig.~1(a). Both the above figures display the particular case of modelling Tri-II by a power law spectrum with $\Gamma =2$. These figures are included here simply for the sake of demonstration.
 
In Table~II, we display the best-fit values of $N_{0}$, along  with the corresponding statistical errors and the test statistics ($\rm{TS}$) values for each of the five values of the spectral indices, employed for the power-spectral modelling of Tri-II, in our analysis. Here, the TS value is defined as~$TS = -2\ln\Big(L_{\rm {(max, 0)}}/L_{\rm {(max, 1)}}\Big)$, in which the square root of TS is approximately the detection significance of a particular source, while $L_{\rm {(max, 0)}}$ is the Maximum Likelihood (ML) value for a model, in which the source (i.e., Tri-II) under study is removed (the so-called `null hypothesis'), and $L_{\rm {(max, 1)}}$ is the corresponding ML value for the full model. In Table~II, we find that, for each of the assumed spectral indices of Tri-II, the corresponding normalisation constant $N_{0}$ is less than the magnitude of the statistical errors involved in the fitting procedure. Also, the corresponding TS value is less than unity. Thus, the Table~II seems to imply that the LAT could detect no signal of any significance in the direction of Tri-II.

As no significant signal has been detected by LAT in the direction of Tri-II, we are required to derive the upper limit of the possible $\gamma$-ray flux from Tri-II. This upper limit is evaluated over the full dataset (i.e., over the entire range of reconstructed energy of photons from $100$~MeV to $50$~GeV) by using the profile likelihood method \cite{bib:bar, bib:rol}. In this method, $N_{0}$ is determined with $95 \%$ C.L. by using a procedure, in which all the three normalisations, namely $N_{0}$ and the  normalisations pertaining to the diffuse and the isotropic background, respectively, are fitted with the LAT-obtained spectrum at each step. This procedure is continued until the difference of the logarithm of the likelihood function reaches the value $1.35$ \cite{bib:abdo} corresponding to an one-sided  $95\%$ C.L. We then apply the Bayesian method, as implemented in the \texttt{Fermi ScienceTools} \cite{bib:abdo}, to obtain a more appropriate value for the upper limit of the $\gamma$-ray flux with $95 \%$ C.L. In Table III, we display the upper limits of the $\gamma$-ray flux for different spectral indices in the entire energy range considered above. In this Table, we find that the upper limit of the gamma-ray flux for $\Gamma = 1$ is about $16$ times lower than the one for $\Gamma = 2.4$. This result is consistent with the ones obtained in Ref.~\cite{bib:abdo} for a number of dSphs (excluding Tri-II) observed by the Fermi-LAT detector in the Milky Way Galaxy.

\begin{table}[!h]
\begin{center}
\caption{Fitted values of the normalisation parameter and TS values for five different spectral indices ($\Gamma$) in the chosen ROI.}
\begin{tabular}{|p{3cm}|p{3cm}|p{3cm}|}
\hline 
\hline
Spectral~Index~($\Gamma$) & $\rm{N_{0} \times 10^{-5}}$  & Test Statistic (TS) value \\
\hline 
$1$  & $(1.41\pm2.75)\times10^{-9}$ & 0.41   \\
\hline 
$1.8$   & $(6.66\pm11.49)\times10^{-8}$ & 0.44   \\
\hline 
$2$  & $(1.06\pm2.41)\times10^{-7}$ & 0.23   \\
\hline 
$2.2$   & $(1.88\pm5.53)\times10^{-7}$ & 0.02   \\
\hline 
$2.4$   & $(1.41\pm2.75)\times10^{-11}$ & $-7.45\times 10^{-8}$   \\
\hline 
\hline
\end{tabular}
\end{center}
\end{table}

\begin{table}[!h]
\caption{Flux upper limits of Tri-II at $95\%$ C.L.}
\begin{center}
\begin{tabular}{|p{3cm}|p{8cm}|}
\hline 
\hline
Spectral~Index~($\Gamma$) & Flux~upper~limits~at~$\rm{95\%~C.L.~(cm^{-2}~s^{-1})}$ \\
\hline 
1 & $8.29\times10^{-11}$ \\
\hline 
1.8 & $4.55\times10^{-10}$ \\
\hline 
2 & $7.14\times10^{-10}$ \\
\hline 
2.2 & $1.04\times10^{-9}$ \\
\hline 
2.4 & $1.37\times10^{-9}$ \\
\hline
 \hline
\end{tabular}
\end{center}
\end{table}

\section{A theoretical framework to estimate the $\gamma$-ray flux from the pair-annihilation of the WIMPs in Tri-II }

\subsection{Modelling of the dark matter density profile of Tri-II}
At a given energy $E$, the differential $\gamma$-ray flux $\phi_{\gamma}(E, \Delta \Omega)$ (in units of photons~${\rm {cm}}^{-2}$~${\rm s}^{-1}$~${\rm {GeV}}^{-1}$) from the annihilations of the WIMPs of mass ${m_{\rm{WIMP}}}$ in a region within a solid angle $\Delta \Omega$ and centred on Tri-II, which is assumed to be a DM source, may be expressed as \cite{bib:bal,bib:abdo,bib:hay}
\begin{equation}
\phi_{\gamma}(E, \Delta \Omega)~ = ~ \Phi^{pp}(E) \times J(\Delta \Omega),
\end{equation}
where, $\Phi^{pp}(E)$ (in units of ${\rm {GeV}}^{-3}~{\rm {cm}}^{3}~{\rm s}^{-1}$) and $J(\Delta \Omega)$ (in units of ${\rm {GeV}}^{2}~{\rm {cm}}^{-5}$) may be denoted as the ``particle physics factor" and the ``astrophysical factor" (or, the``J factor"), respectively. In the following, we discuss these two factors in turn in some more details.

\subsubsection{\textbf{Particle physics factor}}
The factor $\Phi^{pp}(E)$ depends only on the characteristics of the candidate particle of DM. For the WIMPs in particular, the factor may be written as \cite{bib:abdo}
\begin{equation}
\Phi^{pp}(E)~ = ~ \frac{<\sigma v> }{8 \pi ~m^{2}_{\rm{WIMP}}} \sum_{f} \frac{dN_{f}}{dE}(E, m_{{\rm{WIMP}}})~B_{f},\\
\end{equation}
where, $<\sigma v>$ is the average of the product of the relative velocity and the annihilation cross-sections of two annihilating WIMPs and the average is taken over the velocity distributions of those WIMPs \cite{bib:abdo}. In Eq.~(3), $\frac{dN_{f}}{dE}$ is the differential photon spectrum of each possible pair-annihilation final state `$f$', while $B_{f}$ is the branching fraction corresponding to the $f^{\rm{th}}$ final state. The summation in Eq.~(3) runs over all possible final $f$ states. We would like to add that, here, we did not consider the Sommerfeld enhancement \cite{bib:ark,bib:abdo, bib:feng} i.e., the increment in $\gamma$-ray flux due to dependence of annihilation cross-section on relative velocity of particles. This factor comes in as the relative velocity of thermal relics at freeze-out is different than at the present epoch. Hence, the numerical value of annihilation cross-section may differ. Sommerfeld enhancement factor takes into account that mismatch in relative velocity and maximizes the signal by a factor $7$ to $90$ for DM mass in the range of $100$ GeV to $3$ TeV \cite{bib:feng}. In order to be conservative, we did not include such effect.
\subsubsection{\textbf{Astrophysical factor}}

Astrophysical factor or J-factor is related to DM density distribution in Tri-II. The J-factor can be defined as

\begin{equation}
J (\Delta \Omega) = \int \int \rho^{2}(r(\lambda)) d\lambda ~ d\Omega,
\end{equation}

where, $\rho(r)$ is the assumed mass density of DM in Tri-II at the point under observation, situated at a distance $r$ from the centre of Tri-II. In Eq.~(4), we consider $\lambda$ to be the line-of-sight (l.o.s) distance and $r(\lambda)$ to be the galactocentric distance of this observed point in Tri-II. After using the relation $r(\lambda) = \sqrt{\lambda^{2} + d^{2} - 2~ \lambda ~d~ \rm{cos \theta}}$ \cite{bib:evansarkar}, with $d$ being the heliocentric distance and $\theta$ is the angle between the direction of observation and the center of Tri-II, we may rewrite Eq.~(4) as:

\begin{equation}
    J = 2 \pi \int_{0}^{\theta_{\rm{max}}} \rm{sin} \theta \int_{\lambda_{\rm{min}}}^{\lambda_{\rm{max}}} \rho^{2}\Big(\sqrt{\lambda^{2} + d^{2} - 2~ \lambda ~d~ \rm{cos \theta}} \Big) d\lambda ~ d\theta,
\end{equation}

\noindent in which $\theta_{\rm {max}}$ is the angle required to average the expression of the astrophysical factor over the solid angle

\begin{equation}
\Delta \Omega = 2\pi(1-\cos{\theta_{\rm {max}}}).
\end{equation}

\noindent In Eq.~(5), the quantities $\lambda_{\binom{\rm {max}}{\rm {min}}}$ are the lower and the upper limits of the l.o.s integration, that are usually defined as $\lambda_{\binom{\rm {max}}{\rm {min}}} = d~\cos\theta \pm \sqrt{r_{t}^{2} - d^{2}\sin^{2}{\theta}}$~\cite{bib:evansarkar}, respectively with $r_{t}$ being the tidal radius of Tri-II. We also note that, as the LAT observes Tri-II as a point source, the quantity $\Delta \Omega$ for this source is expected to be larger than the resolution ($\theta \lesssim 0.1^{\circ}$ for $E \geq 1$~GeV) of the LAT. In this paper, we have modelled the mass density of the DM distribution of Tri-II with Navarro-Frenk-White (NFW) density profile \cite{bib:nav}. For NFW density profile, we can use the following approximate relation to calculate J-factor \cite{bib:evan1}, i.e.,
\begin{equation}
J \approx\frac{25}{8G^{2}} \frac{\sigma_{{\rm{v}}}^{4}\theta}{dr_h^{2}} .
\end{equation} 
where, $\sigma_{{\rm{v}}}$ is the velocity dispersion, $r_{h}$ is 2D projected half light radius and $G$ is the gravitational constant. We have observed that the estimated values  of J-factor from the above relation are in good agreement with the numerically estimated values of J-factor for different dSphs. For our calculation purpose, we have considered $\theta  = 0.15^{\circ}$ \cite{bib:gen}. In Table IV, we have shown two different J-values corresponding to two different $\sigma_{{\rm{v}}}$.
 
\begin{table}[!h]
\begin{center}
\caption{Different parameters to calculate the astrophysical factor (J), see text for details.}
\begin{tabular}{|p{2cm}|p{2cm}|p{2cm}|p{1cm}|p{3cm}|}
\hline 
\hline
d (kpc) \cite{bib:lae} & $\sigma_{{\rm{v}}}$ ($\rm{km~s^{-1}}$) \cite{bib:kir1} & $r_{h}$ (pc) \cite{bib:kir1}& $\theta$ (deg) \cite{bib:gen}& J-factor from Eq.~(7) ($\rm{{GeV^{2}~cm^{-5}}}$)\\
\hline 

$\rm{30\pm2}$  & $4.2~(95\%~\rm{C.L.})$ & 34 & $0.15^{\circ}$ & $\rm{0.17\times10^{20}}$ \\
\hline
$\rm{30\pm2}$ &  $3.4~(90\%~\rm{C.L.})$ & 34  & $0.15^{\circ}$ & $\rm{0.75\times10^{19}}$ \\
\hline
\hline
\end{tabular}
\end{center}
\end{table}

In our J-factor calculation, we did not take into account the contribution due to the annihilation in cold and dense substructures in Tri-II which in principle can increase the value of J \cite{bib:abdo}. However, we did not include such effect for the present calculation as the previous studies have shown that such effect can boost the J-factor by only a factor of few \cite{bib:mart,bib:abdo}.

\subsection{Constraints on annihilation cross-section}
We have obtained $95 \%$ C.L. upper limits on $\gamma$-ray fluxes (i.e., integral flux above 100 MeV) and on $<\sigma v>$ as a function of the WIMP mass for specific annihilation channels with the help of Eq.~(2), the estimated J-value from Eq.~(7) (see. Table IV) and the DMFit package \cite{bib:jel}  implemented in the \texttt{ScienceTools}. For our analysis, we have chosen pair-annihilation final states such as $b\bar{b}$, $\tau^{+}\tau^{-}$, $\rm{\mu^{+} \mu^{-}}$ and $W^{+}W^{-}$. Such final states, in particular, are highly motivated by the case of neutralino candidates predicted by supersymmetry \cite{bib:jung}. Previously, such annihilation final states have been also used to study the \textit{Fermi} data of dSphs \cite{bib:abdo}. Although, our choice of final states is motivated here by the supersymmetry, the results presented in this section are not restricted only for neutralinos rather they are applicable for generic WIMP models. 
\begin{figure}
\begin{center}
\includegraphics[width=0.6\textwidth,clip,angle=0]{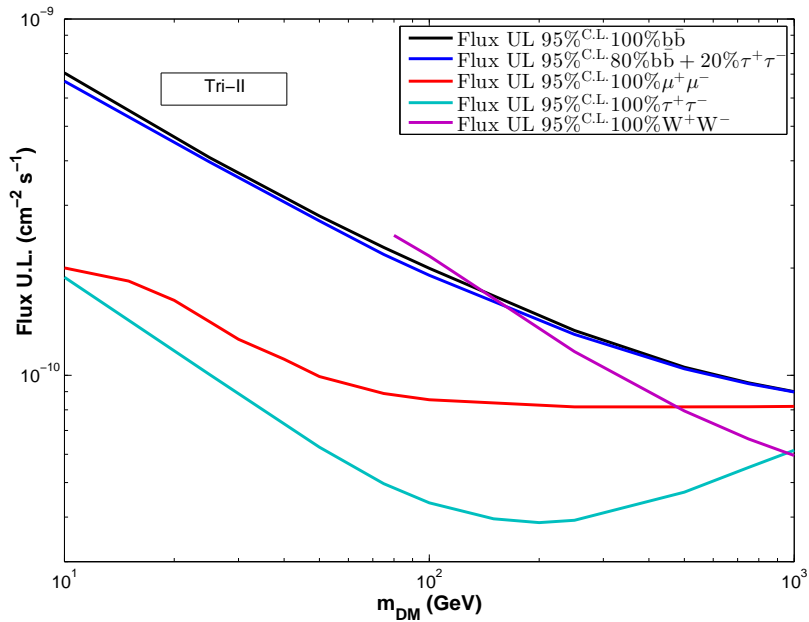}
\caption{Derived upper limits on $\gamma$-ray (integral) fluxes of Tri-II for various annihilation final states are shown in the figure.}
\end{center}
\end{figure}

In Fig.~2, we have shown the variation of the upper limits on the photon fluxes (i.e., integral fluxes above $100$~MeV) corresponding to various annihilation final states, such as $100\%$ b$\rm{\bar{b}}$, $100\%$ $\rm{\tau^{+} \tau^{-}}$, $80\%$ b$\rm{\bar{b}} + 20\%$ $\rm{\tau^{+} \tau^{-}}$, $100\%$ $\rm{\mu^{+} \mu^{-}}$ and $100\%$ $\rm{W^{+} W^{-}}$ (these percentages are also useful for the direct comparison of our results with the previous results of dSphs), with the mass of WIMP. Among those final annihilation states producing hard $\gamma$-ray spectrum, the results for  $\rm{\mu^{+} \mu^{-}}$ and $\rm{\tau^{+} \tau^{-}}$ are considered to be the best upper limits as they predict abundant photon flux at higher energies where the diffuse background is lower. Fig.~2 also shows that for $m_{\rm{DM}} \sim 1$~TeV the upper limits related to all final states vary within a factor of $3$ whereas at lower masses that variation is more than an order of magnitude. The results, as shown in Fig.~2, do not depend on any particular particle theory since they are based only on the final states of WIMP annihilation. We consider now few specific models to study the annihilation cross-section of WIMPs.

\begin{figure}[!h]
\subfigure[]
 { \includegraphics[width=0.65\linewidth]{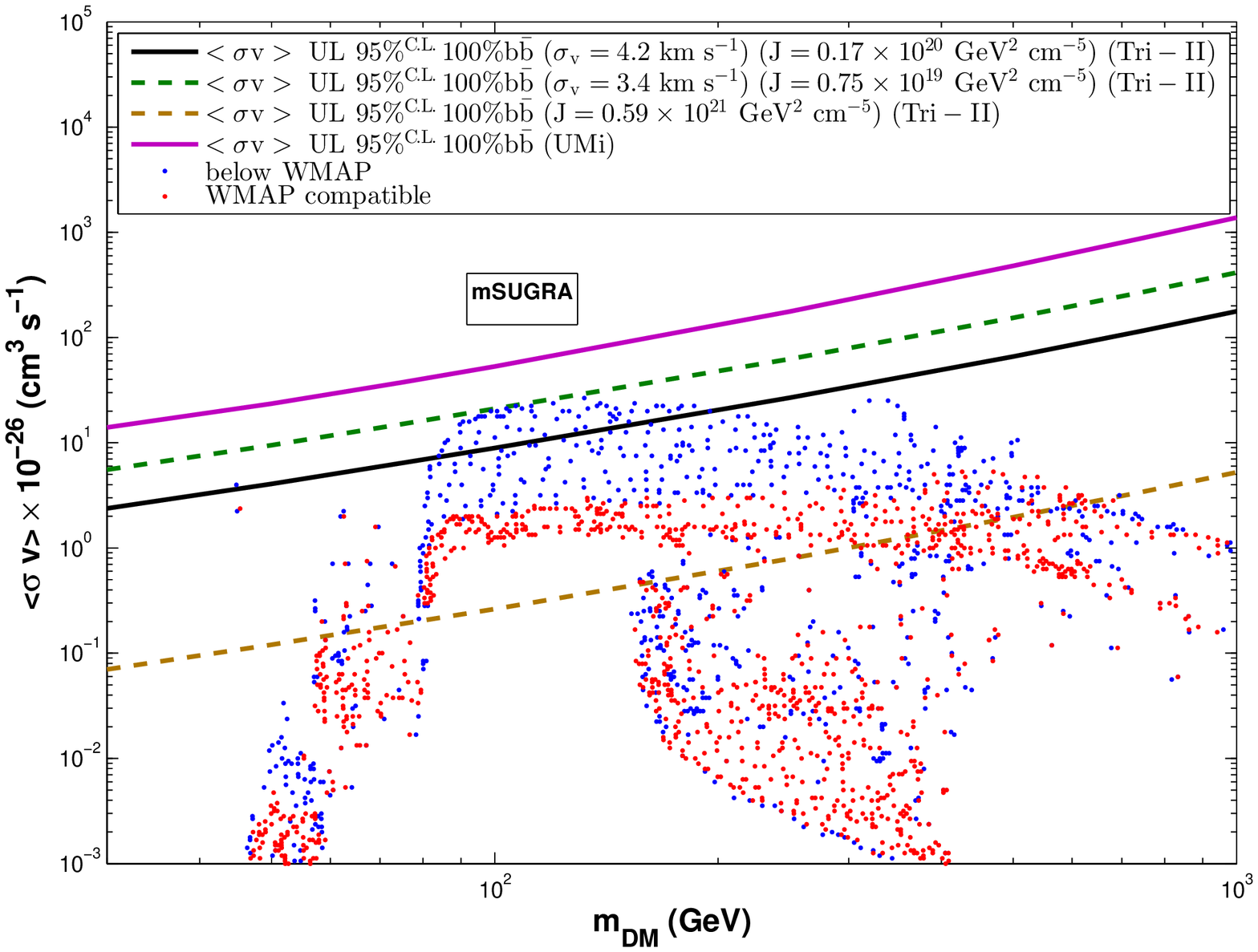}}
\subfigure[]
 { \includegraphics[width=0.65\linewidth]{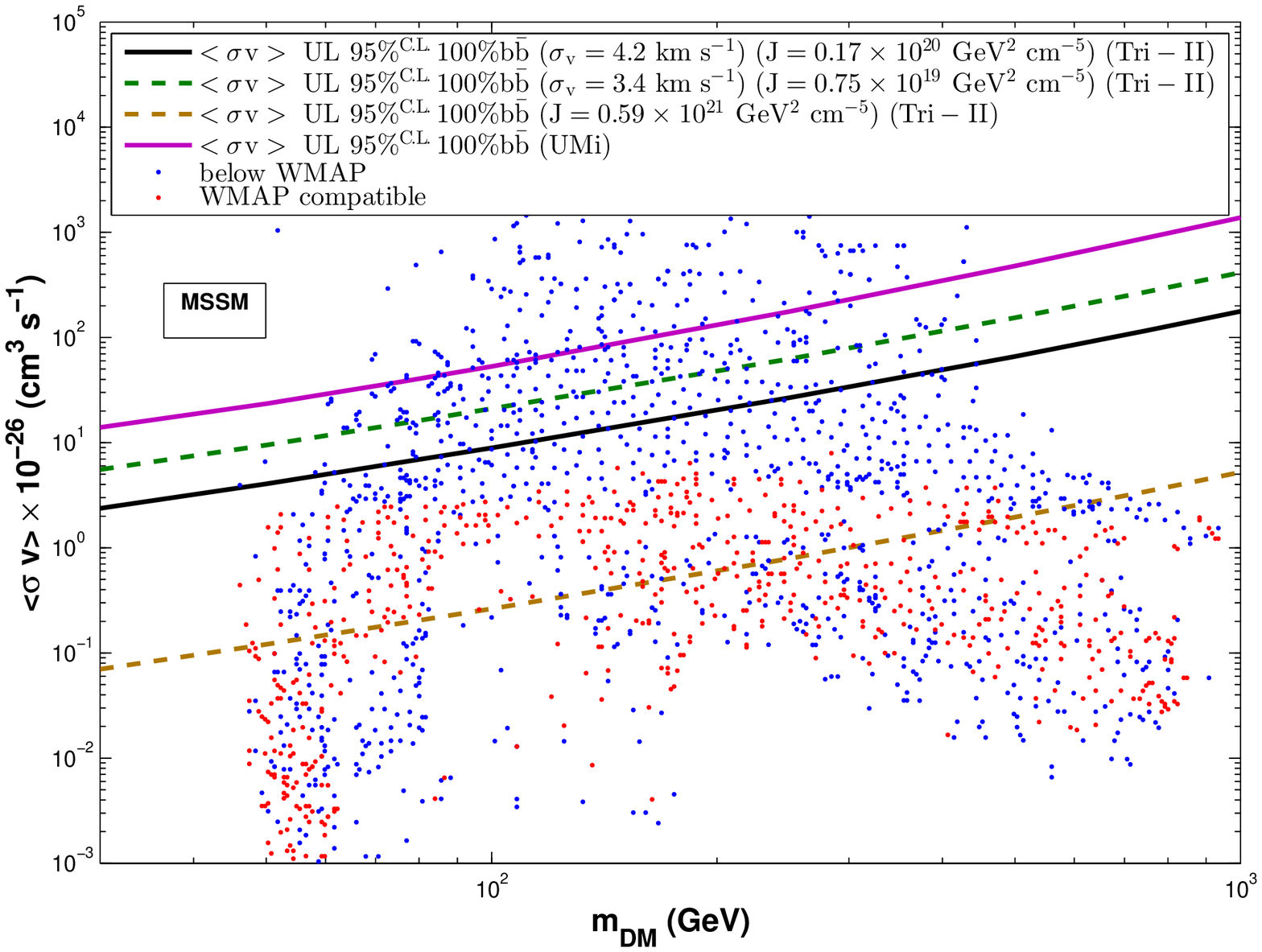}}
 \caption{Predictions from (a) mSUGRA and (b) MSSM models are plotted in ($\rm{m_{WIMP},<\sigma v>}$) plane. The red points (in the two panels) denote neutralino thermal relic abundance related to the inferred cosmological DM density and the blue points correspond to lower thermal relic density. For b$\rm{\bar{b}}$ channel,  the upper limits on $<\sigma~v>$, considering the two velocity dispersion values (\cite{bib:kir1}), for Tri-II have been obtained with $95\%$ C.L. Similarly, the upper limits on $<\sigma~v>$ for  UMi  have also been obtained for the same channel and C.L. The yellow lines (in both the plots) also show the upper limits on $<\sigma v>$, for b$\rm{\bar{b}}$ channel with $95\%$ C.L., for Tri-II with a higher J value which is predicted in Ref.~\cite{bib:gen}. The data of red and blue points are obtained from Ref.~\cite{bib:abdo} and we have produced the data of UMi using the DMFit package and the parameter set used in Ref.~\cite{bib:abdo}.}
 \end{figure}

 \begin{figure}

 {\includegraphics[width=0.6\linewidth]{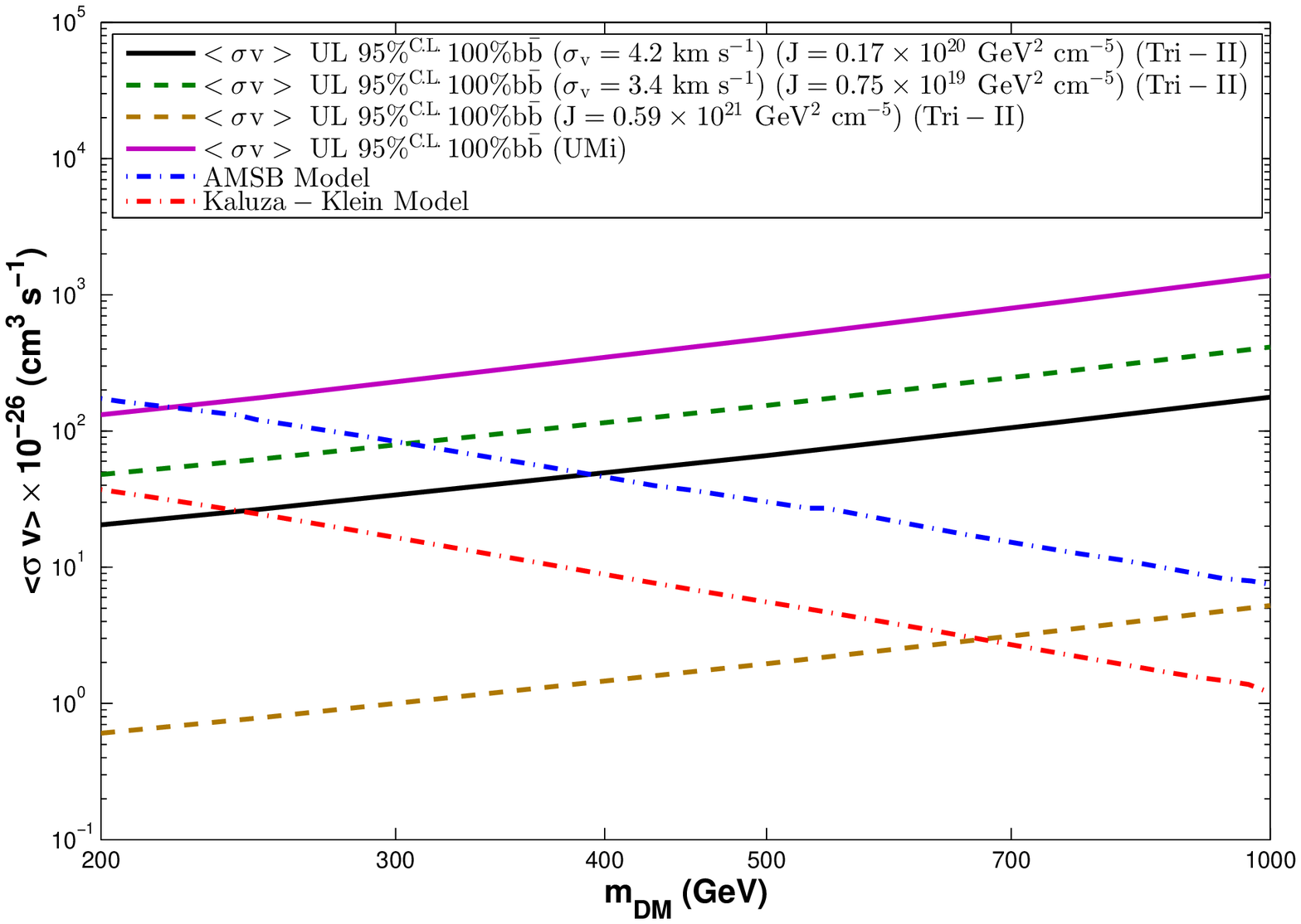}}
\caption{Predictions from AMSB and Kaluza-Klein UED models are plotted in ($\rm{m_{WIMP},<\sigma v>}$) plane. For b$\rm{\bar{b}}$ channel,  the upper limits on $<\sigma~v>$, considering the two velocity dispersion values (\cite{bib:kir1}), for Tri-II have been obtained with $95\%$ C.L. Similarly, the upper limits on $<\sigma~v>$ for  UMi  have also been obtained for the same channel and C.L. The yellow line also shows the upper limit on $<\sigma v>$, for b$\rm{\bar{b}}$ channel with $95\%$ C.L., for Tri-II with a higher J value which is predicted in Ref.~\cite{bib:gen}. The data of Kaluza-Klein UED and AMSB models are obtained from Ref.~\cite{bib:abdo} and we have produced the data of UMi using the DMFit package and the parameter set used in Ref.~\cite{bib:abdo}.}
\end{figure}

In Figs.~3(a,b) and Fig.~4, we have compared the resulting LAT sensitivity for Tri-II for three J values and also for a well known dSph, namely, Ursa Minor (UMi), the latter being produced by us from the Fermi-LAT archival data for an observational period of 11 months, with different predictions from the different theoretical models. In this list first two models are minimal supergravity (mSUGRA) \cite{bib:cham} and Minimal Supersymmetric Standard Model (MSSM) \cite{bib:chu}. mSUGRA is a theoretically motivated DM model. In this model, the supersymmetry breaking parameters are specified at high energy scale which is typically of the order of  grand unification scale $\sim 2 \times 10^{16}$~GeV. On the other hand, all the supersymmetry breaking parameters of MSSM are defined in the electro-weak energy (low energy) scale. Next, we have considered the anomaly mediated supersymmetry breaking (AMSB) model \cite{bib:giu} where the supersymmetry breaking scenario may produce wino-like neutralinos or winos (i.e., a mass eigenstate of neutralino which corresponds to the supersymmetric fermionic partners of the SU(2) gauge bosons of the Standard Model). For about $2$ TeV wino mass, the universal DM density matches with the thermal relic density produced by winos. Several non-thermal production scenarios also exist that could explain the wino DM scenario with lighter (i.e., DM mass is less than a TeV) DM candidates \cite{bib:abdo}. The last one is the lightest Kaluza-Klein particle of universal extra dimensions (UED) \cite{bib:che,bib:ser, bib:hoop}. In this model, in the minimum setup, the first order excitation of the U(1) hypercharge gauge boson which is also commonly known as $\rm{B}^{(1)}$ is related to the DM candidate. Actually in this case, an almost exact relationship exists between DM mass and its pair annihilation cross-section, and a thermal relic abundance related to DM density can be obtained for DM masses around $700$~GeV \cite{bib:ser, bib:abdo}.

In Figs.~3(a,b) and Fig.~4, we have compared the LAT sensitivity, obtained during the analysis of \textit{Fermi} data of Tri-II with DMFit package, in the ($\rm{m_{DM}}$, $<\sigma v>$) plane with the predicted estimates from our four selected DM models namely mSUGRA, MSSM, Kaluza-Klein DM in UED and wino-like DM in AMSB. Red points, in Figs.~3(a,b), are consistent with the $3 \sigma$ WMAP constraint on the universal matter density in accord with thermal production while the blue points would indicate the lower thermal relic density \cite{bib:abdo}. In this paper, we assume that the blue points are related to non-thermal production mechanism of WIMPs in order to explain the observed universal matter density, and this kind of assumption was also taken in Ref.~\cite{bib:abdo}. The advantage of such assumption is that neutralino density needs not to be rescaled which would be the case if we would assume exclusive thermal production scenario. Such non-thermal production scenario of WIMPs has been envisioned by various theories. For an example, several string-theory motivated frameworks predict that the generic decay of moduli produce Standard Model particles along with their supersymmetric partners, and those supersymmetric partners will eventually decay into the lightest neutralinos \cite{bib:mor}. Apart from that, decay of topological objects such as Q-balls may produce neutralinos out of equilibrium \cite{bib:fuji1}. A scenario that predicts the presence of a dynamical ``quintessence'' field in a kinematic-dominated phase \cite{bib:sala,bib:pro2} also supports such non-thermal production of WIMPs.

Figs.~3(a,b) and 4 show the upper limits on $<\sigma v>$ for Tri-II for two different values of velocity dispersion, one an optimistic value of $\sigma_{{\rm{v}}}$ $<$ 4.2~km~s$^{-1}$ and another a rather conservative value of $\sigma_{{\rm{v}}}$ $<$ 3.4~km~s$^{-1}$ as stated in Ref.~ \cite{bib:kir1} along with the predictions from mSUGRA, MSSM, AMSB and Kaluza-Klein UED models respectively. In addition, we also compare our results with UMi, one of the best known candidates of DM. It is seen from Figs.~3(a,b) that even for a low velocity dispersion value of 3.4~km~s$^{-1}$, the constraints obtained on mSUGRA and MSSM models with low thermal densities are almost a factor 2.5 lower than that obtained from UMi for $\rm{m_{WIMP}}= 100$ GeV. The constraints improve to a factor of $\sim$ 6 if one considers a velocity dispersion of 4.2~km~s$^{-1}$. Furthermore, Fig~4 also
indicates that for $\sigma_{{\rm{v}}}$ = 4.2~km~s$^{-1}$ the upper limits on $<\sigma v>$ disfavor the Kaluza-Klein in UED and AMSB models with masses $\lesssim 230$~GeV and $\lesssim 375$~GeV respectively. For $\sigma_{{\rm{v}}}$ = 3.4~km~s$^{-1}$, the AMSB models are disfavored for masses $\lesssim 300$~GeV. However, no effective constraints can be put on Kaluza-Klein in UED models for such a conservative velocity dispersion. It must also be noted that we have only presented the results for $100\%$ b$\rm{\bar{b}}$ channel as it puts stronger constraints on the theoretical models than the other channels considered in the paper.

We here want to point out the fact that for a higher J value (i.e., J~$= \rm{0.59\times10^{21}}~\rm{{GeV^{2}~cm^{-5}}}$) of Tri-II, as predicted in Ref.~\cite{bib:gen}, the constraints on theoretical models as obtained from that J value are more stringent (see Figs.~3(a,b) and 4) than those obtained by us from the J values which are estimated from the velocity dispersion values of Tri-II. For example, at $m_{\rm{DM}}= 100$~GeV, J~$= \rm{0.59\times10^{21}}~\rm{{GeV^{2}~cm^{-5}}}$ predicts the value of $<\sigma~v>$ that is lower by a factor $ \sim 30$ than the obtained $<\sigma~v>$ value from J~$= \rm{0.17\times10^{20}}~\rm{{GeV^{2}~cm^{-5}}}$. Furthermore, such high J value disfavors the Kaluza-Klein in UED for mass $<~700$~GeV and AMSB model for mass range  $<~1000$~GeV.

\begin{figure}

 {\includegraphics[width=0.6\linewidth]{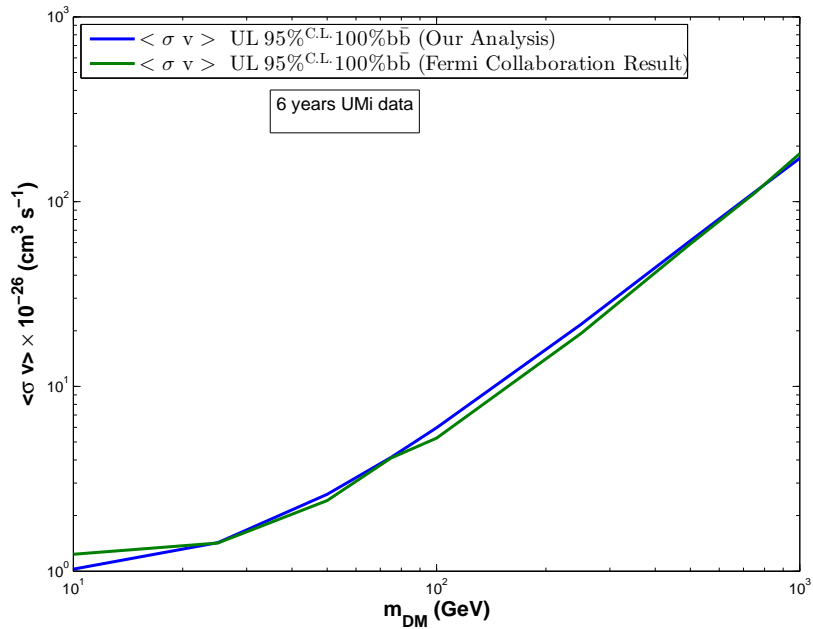}}
\caption{Comparison of constraints on $<\sigma~v>$ for the b$\rm{\bar{b}}$ channel as obtained by us in our analysis with the one obtained by \textit{Fermi} collaboration in the analysis of UMi \cite{bib:ack}.}
\end{figure}

We would also like to add that we have analysed the data of UMi and compared our result with that obtained by \textit{Fermi} collaboration \cite{bib:ack} by applying the same data selection and analysis procedure. The results are shown in Fig 5. 
From the comparison, as shown in Fig.~5, we can conclude that our result closely matches with the result obtained by \textit{Fermi} collaboration. This justifies the reliability of our analysis procedure followed in this paper.

\section{Conclusions}
In this work, we have analysed the $\gamma$-ray data in the direction of Tri-II by \texttt{Fermi ScienceTools}. We do not observe any excess $\gamma$-ray emission from Tri-II and upper limit has been derived on the $\gamma$-ray flux from Tri-II.

Using the DM halo modelling, we have further estimated the upper limits of photon flux and $<\sigma v>$ by considering that the DM entirely consists of neutralinos. Our results show that in Tri-II, for $\rm{\sigma_{v}= 4.2~km~s^{-1}}$, $100\%$ b$\rm{\bar{b}}$ channel constrains mSUGRA and MSSM models with low thermal relic densities and Kaluza-Klein DM in UED and AMSB models with masses $\lesssim 230$~GeV and $\lesssim 375$~GeV respectively. It is important to note for the velocity dispersion with 90 $\%$ C.L. i.e., $\rm{\sigma_{v} = 3.4~km~s^{-1}}$ also constrains the MSSM model with low thermal relic densities and AMSB model with masses $\lesssim 300$~GeV. In addition, we showed that more higher J value can put more stronger limits on parameter space of  existing theoretical models of DM. Furthermore, we would like to add that the analysis of $\gamma$-ray data from Tri-II puts stronger limits on the DM models than UMi which justifies our choice of Tri-II for testing the DM models.  We would like to mention another point that the results are based on the standard NFW halo shape and we do not take into account the effects of boost factor related to substructures in Tri-II or the Sommerfeld effect in accord to the annihilation cross-section. Finally, we can say that more precise observations of Tri-II in future would ultimately vindicate the possibility of establishing Tri-II as a candidate for indirect DM search.

\section*{Acknowledgement}
We thank the anonymous referee for pointing out some valuable points which improve the paper a lot. PB is grateful to the DST INSPIRE FELLOWSHIP for financial support. We are grateful to Dr. Stefan Profumo and Dr. Tesla Jeltema of the University of California, Santa Cruz, USA for much needed help, advice and guidance in the Fermi-LAT Data analysis. We would also like to thank Dr. James Unwin, Visiting Research Assistant Professor, University of Illinois at Chicago, IL, USA, for providing useful feedback which helps us to improve the present manuscript. We are grateful to the Fermi Science Support Center for providing free access to the Fermi data and an excellent guidance for analysing those data in the form of the Fermi Science Tool manuals.

\end{document}